\documentclass[twocolumn,showpacs,preprintnumbers,amsmath,amssymb,prl]{revtex4}

\usepackage{graphicx}
\usepackage{dcolumn}
\usepackage{bm}
\usepackage{amssymb}

\begin{document}

\title{Dissipationless Electron Transport in Photon-Dressed Nanostructures}

\author{O.V. Kibis}\email{Oleg.Kibis@nstu.ru}

\affiliation{Department of Applied and Theoretical Physics,
Novosibirsk State Technical University, Karl Marx Avenue 20,
630092 Novosibirsk, Russia}


\begin{abstract}
It is shown that the electron coupling to photons in field-dressed
nanostructures can result in the ground electron-photon state with
a nonzero electric current. Since the current is associated with
the ground state, it flows without the Joule heating of the
nanostructure and is nondissipative. Such a dissipationless
electron transport can be realized in strongly coupled
electron-photon systems with the broken time-reversal symmetry ---
particularly, in quantum rings and chiral nanostructures dressed
by circularly polarized photons.
\end{abstract}

\pacs{73.63.Nm, 42.50.Ct, 74.20.Mn}

\maketitle

The scientific trend, that has emerged in recent years, is to
combine the methods of quantum optics with the advances in design
and preparation of various nanostructures. This interdisciplinary
and innovative field of physics, arisen from achievements of the
modern nanotechnology, offers the possibility of both the
fundamental studies of light-matter interaction in unusual
artificial quantum systems and the development of optoelectronic
nanodevices with unique characteristics (see, e.g.,
Ref.~\cite{Kibis_09}). In the present Letter, the methodology of
quantum optics is applied to analyze the strong interaction
between electrons in curvilinear nanostructures and a quantized
electromagnetic field. It follows from the analysis that the
electron coupling to photons can result in the ground
electron-photon state with a nonzero electric current. Since the
current is associated to the ground state, it flows without the
Joule heating of the nanostructure and is nondissipative. Thus, we
declare a novel mechanism of the photon-induced superconductivity.
The main goal of the Letter is to present the first theoretical
analysis of this mechanism.

Let us consider the effect of an electromagnetic field on electron
transport in curvilinear one-dimensional conductors (quantum
wires) of a constant cross section and of a constant curvature ---
specifically, in ring-shaped and helix-shaped quantum wires
(quantum rings and quantum helices, respectively). In what
follows, the spin effects and the field-induced mixing of
transverse electron modes in the quantum wire will be beyond
consideration. Then, the processes of electron-field interaction,
associated to the electron motion along the wire, can be described
by the Hamiltonian $\hat{\mathcal H}_e=\hat{p}_s^2/2m_e+ U(s)$,
where $s$ is the electron coordinate along the wire,
$\hat{p}_s=-i\hbar\partial/\partial s$ is the operator of electron
momentum along the wire, $U(s)=-e\int E_t(s)ds$ is the potential
energy of an electron in the wire subjected to the electromagnetic
field, $e$ is the electron charge, $m_e$ is the electron mass, and
$E_t(s)=\mathbf{E}\mathbf{t}(s)$ is the projection of the electric
field $\mathbf{E}$ on a tangent to the quantum wire with the unit
vector $\mathbf{t}(s)$. As to the magnetic component of the
electromagnetic field, it does not influence the electron motion
along the quantum wire since the Lorentz force is perpendicular to
the wire. Considering the problem within the conventional
quantum-field approach \cite{Landau_4}, the classical field
$\mathbf{E}$ should be replaced with the field operator
$\hat{\mathbf{E}}$. Then the full Hamiltonian of the
electron-photon system, including both the field energy
$\hbar\omega\hat{a}^\dagger\hat{a}$ and the electron Hamiltonian
$\hat{\mathcal H}_e$ is
\begin{equation}\label{H}
\hat{\mathcal
H}=\hbar\omega\hat{a}^\dagger\hat{a}+\frac{\hat{p}_s^2}{2m_e}-e\int\hat{\mathbf{E}}\mathbf{t}(s)ds\;,
\end{equation}
where $\hat{a}$ and $\hat{a}^\dagger$ are the operators of photon
annihilation and creation, respectively, written in the
Schr\"{o}dinger representation (the representation of occupation
numbers) \cite{Landau_4}, and $\omega$ is the frequency of the
electromagnetic field.  For definiteness, in this Letter we will
discuss nanostructures exposed to a plane monochromatic circularly
polarized electromagnetic wave. Assuming the electromagnetic wave
to be clockwise polarized and neglecting any spatial inhomogeneity
of the wave field, the field operator in Eq.~(\ref{H}) can be
written as
$\hat{\mathbf{E}}=i\sqrt{2\pi\hbar\omega/V}\left(\mathbf{e}_+\hat{a}-
\mathbf{e}_-\hat{a}^\dagger\right)$, where $V$ is the quantization
volume, $\mathbf{e}_{\pm}=(\mathbf{e}_{x}\pm
i\mathbf{e}_{y})/\sqrt{2}$ are the polarization vectors,
$\mathbf{e}_{x,y,z}$ are the unit vectors directed along the
$x,y,z$ axes of the Cartesian coordinate system, and the $z$ axis
corresponds to the direction of the wave propagation
\cite{Landau_4}.

First of all, let us consider a ring-shaped quantum wire (a
quantum ring with the radius $R$) subjected to a circularly
polarized electromagnetic wave propagating along the ring axis.
Since the tangent vector to the ring is
$\mathbf{t}(s)=-\sin\left(s/R\right)\mathbf{e}_x+\cos\left(s/R\right)\mathbf{e}_y$,
the Hamiltonian (\ref{H}) takes the form $\hat{\mathcal
H}=\hat{\mathcal H}^{(0)}_{R}+\hat{U}_R$, where
\begin{equation}\label{UR}
\hat{U}_R=-ieR\sqrt{\frac{\pi\hbar\omega}{V}}\left(e^{i\varphi}\hat{a}-e^{-i\varphi}\hat{a}^\dagger\right)
\end{equation}
is the Hamiltonian of the electron-photon interaction in the ring,
$\hat{\mathcal
H}^{(0)}_{R}=\hbar\omega\hat{a}^\dagger\hat{a}+{\hbar^2\hat{l}_z^2}/{2m_eR^2}$
is the Hamiltonian of a noninteracting electron-photon system,
$\hat{l}_z=-i\partial/\partial\varphi$ is the operator of the
electron angular momentum along the ring axis, and $\varphi=s/R$
is the angular coordinate in the ring.  To describe the
electron-photon system, let us use the notation $|m,N\rangle$
which indicates that the electromagnetic field is in a quantum
state with the photon occupation number $N=1,2,3,...\;$, and the
electron is in a quantum state with the wave function
$\psi_m(\varphi)=\sqrt{1/2\pi}\exp(im\varphi)$, where
$m=0,\pm1,\pm2,...$ is the electron angular momentum along the
ring axis. The electron-photon states $|m,N\rangle$ are true
eigenstates of the unperturbed Hamiltonian $\hat{\mathcal
H}^{(0)}_{R}$ and their energy spectrum is
$\varepsilon^{(0)}_{m,N}=N\hbar\omega+\hbar^2m^2/2m_eR^2$. In
order to find the energy spectrum of the full electron-photon
Hamiltonian (\ref{H}), let us use the conventional perturbation
theory, considering the term (\ref{UR}) as a perturbation with the
matrix elements
\begin{eqnarray}
\langle
m^\prime,N^\prime|\hat{U}_R|m,N\rangle=-ieR\sqrt{\frac{\pi\hbar\omega}{V}}
\left[\sqrt{N}\delta_{m,m^\prime-1}\delta_{N,N^\prime+1}\right.\nonumber\\
-\left.\sqrt{N+1}\delta_{m,m^\prime+1}\delta_{N,N^\prime-1}\right].\;\;\;\;\;\;\;\;\;\;\;\;\;\;\nonumber
\end{eqnarray}
Performing trivial calculations within the second order of the
perturbation theory, we arrive at the sought energy spectrum of
the Hamiltonian (\ref{H}),
\begin{eqnarray}\label{ES}
\varepsilon_{m,N}&=&\varepsilon^{(0)}_{m,N}+\frac{|\langle
m+1,N-1|\hat{U}_R|m,N\rangle|^2}{\varepsilon^{(0)}_{m,N}-\varepsilon^{(0)}_{m+1,N-1}}\nonumber\\
&+&\frac{|\langle
m-1,N+1|\hat{U}_R|m,N\rangle|^2}{\varepsilon^{(0)}_{m,N}-\varepsilon^{(0)}_{m-1,N+1}}\;.
\end{eqnarray}
Let us assume hereafter that the dressing electromagnetic field
cannot be absorbed by electrons. In order to neglect the
collisional absorption of the field, the condition
\begin{equation}\label{ot}
\omega\tau\gg1
\end{equation}
is assumed to be satisfied, where $\tau$ is the effective electron
life time restricted by a scattering of the electron from all
defects
--- both geometrical and structural --- of the nanostructure. To avoid the optical absorption,
the field frequency $\omega$ is assumed to be far from the
resonant frequencies of the ring. Then the allowed energies of the
complete electron-photon system are described by Eq.~(\ref{ES})
for $N=N_0$, where $N_0$ is the constant photon occupation number
of the field. If the electromagnetic wave is strong ($N_0\gg1$),
the energy spectrum (\ref{ES}) can be written as
$\varepsilon_{m,N_0}=N_0\hbar\omega+\varepsilon(m)$, where
$N_0\hbar\omega$ is the constant field energy, the electron-photon
term is
\begin{equation}\label{QP}
\varepsilon(m)=m^2\varepsilon_R+
eE_0R\left[\frac{eE_0R/2\varepsilon_R}{(2m-\hbar\omega/\varepsilon_R)^2-1}\right]\,,
\end{equation}
$\varepsilon_R=\hbar^2/2m_eR^2$ is the characteristic electron
energy in the ring, $E_0=\sqrt{4\pi N_0\hbar\omega/V}$ is the
classical amplitude of electric field of the electromagnetic wave,
and the factor in the square brackets is assumed to be much
smaller than unity. As a consequence, we can consider electrons in
the photon-dressed ring as quasiparticles with the energy spectrum
(\ref{QP}) which is shown schematically in Fig.~1(a) for the case
of $\hbar\omega/\varepsilon_R<1$. Let us restrict the analysis of
the electric currents in the ring by the principal (zeroth) order
of the perturbation theory. Then the current of a field-dressed
electron with an angular momentum $m$ is approximately equal to
the current of a free electron with the same angular momentum,
$j_m=me\hbar/2\pi R^2m_e$ \cite{m0}. As a result, the full
electric current in the ring is $j=\sum_{{m}}j_{{m}}$, where the
summation is performed over filled states (\ref{QP}). It is easy
to show that the field-induced splitting of the states (\ref{QP})
with angular momenta $m$ and $-m$ can lead to the nondissipative
current $j$. As an example, the ground state of four field-dressed
electrons corresponds to the filled states (\ref{QP}) with $m=0$
and $m=-1$ [see Fig.~1(a)]. Correspondingly, the full current in
the ring is $j=2j_0+2j_{-1}=-e\hbar/\pi R^2m_e$. Assuming the
radius of quantum ring $R$ to be of the nanometer scale, we obtain
$j\sim10^{-6}$A. Since this current corresponds to the ground
state, it is nondissipative. To clarify the physical nature of the
current, it should be reminded that the equality
$\varepsilon(m)=\varepsilon(-m)$ takes place in the absence of the
field ($E_0=0$) due to the time-reversal symmetry (the Kramers
theorem). Since the time reversal turns clockwise polarized
photons into counterclockwise polarized ones and vice versa, the
electron coupling to the circularly polarized field breaks this
symmetry. Therefore, field-dressed electron states (\ref{QP}) with
angular momenta $m$ and $-m$ are split, which results in the
discussed effect. Since this splitting decreases with increasing
$m$, the effect can be observable in rings whose Fermi energy is
sufficiently low (e.g., semiconductor rings).

Though the discussed nondissipative current differs substantially
from the known persistent currents arising in quantum rings due to
the Aharonov-Bohm effect \cite{Lorke_00}, it still cannot be
identified with the superconductivity declared in the preamble of
the Letter. Indeed, this current flows in a microscopical ring and
is devoid of such a character of superconductivity as a
nondissipative translational motion of electrons. That is why
persistent currents in quantum rings --- regardless of their
physical nature --- are phenomena of purely academic interest.
However, we will demonstrate hereafter that the same
electron-photon coupling leads also to nondissipative electric
currents in macroscopically long conductors. This argument is
crucial to consider the discussed effect as a conceptually novel
mechanism of superconductivity. Among the variety of different
nanostructures, macroscopically long conductors with chiral
symmetry --- helicoidal quantum wires, chiral carbon nanotubes,
DNA-based nanostructures, etc. --- seem prospective for the
realization of such a superconductivity. To substantiate this
statement, we will discuss the simplest chiral nanostructure
--- a quantum helix (helix-shaped quantum wire). Such helices can
be fabricated by methods of the modern nanotechnology \cite{Prinz}
and promise a broad range of interesting phenomena
\cite{Kibis_92,Kibis_95,Kibis_05,Entin}.

Let us consider a right-hand quantum helix with a radius $R$, a
pitch $P$, and a macroscopically large length $L$. Then the
tangent vector to the helix is
$\mathbf{t}(s)=-gR\sin(gs)\mathbf{e}_x+gR\cos(gs)\mathbf{e}_y+(P/l)\mathbf{e}_z$,
where $s$ is the coordinate along the helix, $l=\sqrt{(2\pi R)^2+
P^2}$ is the length of the helix turn, and $g=2\pi/l$. Assuming a
circularly polarized electromagnetic wave to be propagating along
the axis of the helix, we can write the Hamiltonian (\ref{H}) as
$\hat{\mathcal H}=\hat{\mathcal H}^{(0)}_{H}+\hat{U}_H$, where
\begin{equation}\label{UH}
\hat{U}_H=-ieR\sqrt{\frac{\pi\hbar\omega}{V}}\left(e^{igs}\hat{a}-e^{-igs}\hat{a}^\dagger\right)
\end{equation}
is the Hamiltonian of the electron-photon interaction in the
helix, and $\hat{\mathcal
H}^{(0)}_{H}=\hbar\omega\hat{a}^\dagger\hat{a}+\hat{p}_s^2/2m_e$
is the Hamiltonian of a noninteracting electron-photon system. In
order to neglect a disturbance of the field by the helix, the
field frequency, $\omega$, is assumed to be within the
transparency region of the helix \cite{infrared}. To describe the
electron-photon system, let us use the notation $|k,N\rangle$
which indicates that the electromagnetic field is in a quantum
state with the photon occupation number $N=1,2,3,...\;$ and the
electron is in a quantum state with the wave function
$\psi_k(s)=\sqrt{1/L}\exp(iks)$, where $k$ is the electron wave
vector along the helix. The electron-photon states $|k,N\rangle$
are true eigenstates of the unperturbed Hamiltonian $\hat{\mathcal
H}^{(0)}_{H}$ and their energy spectrum is
$\varepsilon^{(0)}_{k,N}=N\hbar\omega+\hbar^2k^2/2m_e$. Therefore,
they form the complete basis of the electron-photon system, which
allows to seek the eigenstates of the full electron-photon
Hamiltonian (\ref{H}) as an expansion
$\sum_{k^\prime,N^\prime}b_{k^\prime,N^\prime}|k^\prime,N^\prime\rangle$,
where the coefficients $b_{k^\prime,N^\prime}$ are some constants.
Substituting this expansion into the Schr\"{o}dinger equation with
the Hamiltonian (\ref{H}) and taking into account that the matrix
elements of the electron-photon interaction (\ref{UH}) are
\begin{eqnarray}\label{MUH}
\langle
k^\prime,N^\prime|\hat{U}_H|k,N\rangle=-ieR\sqrt{\frac{\pi\hbar\omega}{V}}
\left[\sqrt{N}\delta_{k,k^\prime-g}\delta_{N,N^\prime+1}\right.\nonumber\\
-\left.\sqrt{N+1}\delta_{k,k^\prime+g}\delta_{N,N^\prime-1}\right],\;\;\;\;\;\;\;\;\;\;\;\;\;\;
\end{eqnarray}
we arrive at the system of linear algebraic equations
\begin{eqnarray}\label{SH}
\left(\varepsilon^{(0)}_{k,N}-\varepsilon_{k,N}\right)b_{k,N}+\langle
k,N|\hat{U}_H|k+g,N-1\rangle b_{k+g,N-1}\nonumber\\
+\langle k,N|\hat{U}_H|k-g,N+1\rangle b_{k-g,N+1}=0\;,\;\;\;\;\;
\end{eqnarray}
where $\varepsilon_{k,N}$ is the sought energy spectrum of the
Hamiltonian (\ref{H}). The equations (\ref{SH}) under the
condition (\ref{ot}) describe the allowed energies of the
electron-photon system $\varepsilon_{k,N}$ for the constant photon
occupation number $N=N_0$. It follows from Eq.~(\ref{MUH}) that
the electron-photon interaction (\ref{UH}) mixes a set of
electron-photon states with wave vectors $k$ differing by $g$. If
the characteristic energy of the electron-field interaction
$eE_0R$ is much smaller than the characteristic energy interval
between these states $\hbar^2g^2/2m_e$, then it is sufficient to
take into account the mixing of only two states closest in energy,
neglecting the contribution of all other states. In this
approximation the system of equations (\ref{SH}) is reduced to
just two equations. Solving these two equations accurately for
$N_0\gg1$, we obtain the energy spectrum of the electron-photon
system, $\varepsilon_{k,N_0}=N_0\hbar\omega+\varepsilon(k)$, where
the energy of photon-dressed electron is
\begin{equation}\label{ESH}
\varepsilon(k)=\left\{\begin{array}{rl}
\varepsilon_+(k)-\sqrt{\varepsilon^2_-(k)+(\Delta\varepsilon/2)^2}\,
&|k-k_\omega|<g/2\\
\varepsilon_+(k)+\sqrt{\varepsilon^2_-(k)+(\Delta\varepsilon/2)^2}\,
&|k-k_\omega|>g/2
\end{array}\right.,
\end{equation}
$\varepsilon_\pm(k)=({\hbar^2}/{4m_e})\left(k^2\pm k^2\pm
g^2\mp2g|k-k_\omega|\right)$, $k_\omega=m_e\omega/g\hbar$, and
$\Delta\varepsilon=|e|E_0R$. As a consequence, we can consider
electrons in the field-dressed helix as quasiparticles with the
energy spectrum (\ref{ESH}) which is shown schematically in
Fig.~1(b). Correspondingly, the electric current of the
quasiparticle with the wave vector $k$ is $j_k=ev_{k}/L$, where
$v_k=(1/\hbar)d\varepsilon(k)/dk$ is the quasiparticle velocity
along the helix. Therefore the full current in the helix is
$j=\sum_{{k}} j_{{k}}$, where the summation should be performed
over filled states (\ref{ESH}). As an example, the current of a
field-dressed electron system with the Fermi energy $\mu$ [see
Fig.~1(b)] is $j=e\int v_k\;dk/\pi=e\Delta\varepsilon/\pi\hbar$,
where the integration is performed over the filled states lying
under the Fermi level. Since this current is associated to the
ground state of the macroscopically long photon-dressed quantum
wire, it is nondissipative and corresponds to the declared
photon-induced superconductivity.

\begin{figure}[th]
\includegraphics[width=0.40\textwidth]{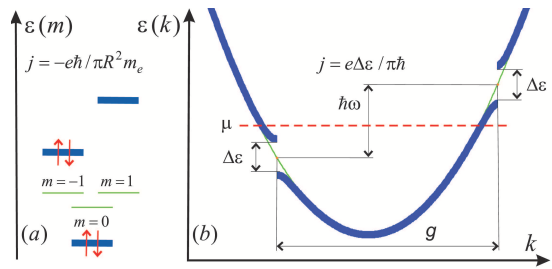}
\caption{(color online) The energy spectrum of electrons --- in
the quantum ring (a) and in the quantum helix (b)
---  dressed by the quantized circularly polarized field (heavy lines)
and without the field (thin lines). The nondissipative currents
$j$ correspond to the filled states ($\uparrow\downarrow$) in the
ring (a) and to the filled states lying under the Fermi energy
$\mu$ in the helix (b).}
\end{figure}

It is seen in Fig.~1(b) that the distinctive feature of the
spectrum (\ref{ESH}) consists in the energy gaps,
$\Delta\varepsilon=|e|E_0R$, positioned asymmetrically in the $k$
space with respect to the band edge --- at $k=k_\omega\pm g/2$.
Just this  asymmetry of the energy gaps leads to the
above-mentioned nondissipative electric current,
$j=e\Delta\varepsilon/\pi\hbar$. Let us clarify the physical
nature of the gaps. It is evident that a stationary electric field
$E_\perp$ oriented perpendicular to the helix axis (a transverse
field) results in a periodic electron potential energy along the
helix, $U(s)=eE_\perp R\cos(gs)$, where the period of this
potential is equal to the length of the helix turn, $l=2\pi/g$. As
a consequence, the Bragg scattering of electrons from such a
periodic potential opens energy gaps at $k=\pm ng/2$
($n=1,2,3,...$) in the electron energy spectra of various
helicoidal nanostructures \cite{Kibis_05,Kibis_05_1}. In contrast
to the stationary case, the rotating transverse electric field of
the circularly polarized electromagnetic wave produces the
periodic time-dependent potential, $U(s,t)=eE_0 R\cos(gs-\omega
t)$, which runs  along the helix with the velocity $v=\omega/g$.
Correspondingly, the Bragg scattering of electrons from this
running periodic potential takes place at $k^\prime=\pm ng/2$,
where $k^\prime=k-k_\omega$ is the electron wave vector in the
reference system moving along the helix with the same velocity $v$
(i.e., in the rest frame of the running potential). As a result,
the Bragg gaps in the energy spectrum (\ref{ESH}) lie at the wave
vectors, $k=k_\omega\pm g/2$, which correspond to the first
diffraction maximum of the Bragg scattering ($n=1$) and are
positioned  asymmetrically in the $k$ space of the laboratory
reference system. As expected, in the limiting case of a
stationary field ($\omega=0$) this asymmetry vanishes
($k_\omega=0$) and the spectrum (\ref{ESH}) exactly coincides with
the electron energy spectrum of the quantum helix exposed to a
stationary transverse electric field, which is given by Eq.~(7) in
Ref.~\cite{Kibis_05}. Certainly, the Bragg gaps also take place
for $n=2,3,4,...$. However, these gaps are lying higher in energy,
depend on higher powers of the electric field, $E_0$, and cannot
be derived from the approximate expression (\ref{ESH}). We will
discuss them elsewhere.

It follows from the aforesaid that the Bragg scattering transfers
electrons between the two states differing by the wave vector $g$
and the energy $\hbar\omega$ [see Fig.~1(b)]. It is also known
that the conservation laws allow optical transitions of free
electrons in the helix between the same two states
\cite{Kibis_95}. Therefore, the Bragg scattering can be
reformulated in terms of the quantum optics: It is identical
physically to the wave-induced Rabi oscillations of electrons
between the mentioned two states with the Rabi frequency
$\Omega=\Delta\varepsilon/\hbar$. Since collisional processes can
erode any energy gap, the condition for the existence of the
discussed gaps depends on the electron lifetime, $\tau$, and is
\begin{equation}\label{oth}
\Omega\tau\gg1\;.
\end{equation}
As expected, the inequality (\ref{oth}) coincides with the
existence condition of the Rabi oscillations \cite{Scully_b01}.
Since the Rabi oscillations are not accompanied by any absorption
of the field energy, the inequality (\ref{oth}) can be interpreted
as the forbidding of resonant field absorption at the wave
frequency $\omega$. Therefore the inequalities (\ref{ot}) and
(\ref{oth}) describe the purely dressing (nonabsorbable) field.
Assuming the helix radius, $R$, to be of the nanometer scale, the
energy gap is $\Delta\varepsilon\sim10^{-2}$~eV for the field
$E_0\sim10^5$~V/cm. In this case the nondissipative current is
$j\sim10^{-6}$A.

It is crucial that the spectra (\ref{QP}) and (\ref{ESH}), which
result in nondissipative currents, are derived for a nonabsorbable
field satisfying the conditions (\ref{ot}) and (\ref{oth}).
Indeed, the absorption of the field energy by electrons leads to
the usual photovoltaic (photon drag) effects in both quantum rings
\cite{Rings} and quantum helices \cite{Kibis_95,Entin}. Since such
a field absorption is accompanied by the energy transfer from the
field to electrons, photovoltaic currents result in the Joule
heating and obey the Ohm law. Therefore, a dressing
electromagnetic field producing nondissipative currents cannot be
absorbable. We have demonstrated that the electron coupling to
such a dressing field forms the physical mechanism of
dissipationless electron transport, which differs substantially
from the traditional mechanisms of the superconductivity.
Generalizing the aforesaid, we can declare that any quantized
periodic potential (a potential wave) running along a conductor
can produce a nondissipative electric current, if the potential
wave is both strong and of sufficiently high frequency. Formally,
the quantized potential wave can be considered as a quantum
version of the electron pump \cite{Thouless_83}. The search for
efficient ways to realize such a quantized pump in various
conducting systems is the subject of future studies.

The work was partially supported by the RFBR (Grants No.
10-02-00077 and No. 10-02-90001), the Russian Ministry of
Education and Science, the 7th European Framework Programme (Grant
No. FP7-230778), and ISTC Project No. B-1708.


\begin{thebibliography}{99}
\bibitem{m0}
In higher orders of the perturbation theory, a nonzero current is
associated also to the state with $m=0$.
\bibitem{infrared}
If the helix is fabricated with using semiconductor materials
\cite{Prinz}, such a field frequency lies in the infrared range.
\end{thebibliography}
\end{document}